\documentclass[12pt]{iopart}

\usepackage{iopams}
\usepackage{cite}
\usepackage{cleveref}

\newtheorem{proposition}{Proposition}
\newtheorem{lemma}{Lemma}
\newtheorem{corollary}{Corollary}

\usepackage[normalem]{ulem}

\newcommand{\cpn}{$\mathbb{C}P^{N-1}$}
\newcommand{\nn}{\nonumber}
\begin{document}

\title[On a stack of surfaces obtained from the $\mathbb{C}P^{N-1}$ sigma models]{On a stack of surfaces obtained from the $\mathbb{C}P^{N-1}$ sigma
models}

\author{P P Goldstein$^1$ and A M Grundland$^{2,3}$}

\address{$^1$ Theoretical Physics Division, National Centre for Nuclear Research,\\ Ho\.za 69, 00-681 Warsaw Poland}
\address{$^2$ Department of Mathematics and Computer Science, Universit\'e du Qu\'ebec, Trois-Rivi\`eres, CP 500 (QC) G9A 5H7, Canada}
\address{$^3$ Centre de Recherches Math\'ematiques, Universit\'e de Montr\'eal,\\ Montr\'eal CP 6128 Succ. Centre-Ville (QC) H3C 3J7, Canada}
\ead{Piotr.Goldstein@ncbj.gov.pl and grundlan@crm.umontreal.ca}
\vspace{10pt}
\begin{indented}
\item[]\today
\end{indented}

\begin{abstract}
Under the assumption that the $\mathbb{C}P^{N-1}$ sigma model is
defined on the Riemann sphere and its action functional is finite,
we derive surfaces induced by surfaces and we demonstrate that the
stacked surfaces coincide with each other, which means idempotency
of the recurrent procedure. Along the path to the solutions of the
$\mathbb{C}P^{N-1}$ model equations, we demonstrate that the
Euler--Lagrange equations for the projectors admit larger classes
of solutions than the ones corresponding to rank-1 projectors.
\end{abstract}


\vspace{2pc} \noindent{\it Keywords}: sigma models, projector
analysis, soliton surfaces, integrable systems.

\submitto{\JPA}

\maketitle

\section{Introduction}
Soliton surfaces associated with integrable models and with the
$\mathbb{C}P^{N-1}$ sigma model in particular have been shown to
play an essential role in many problems with physical applications
(see e.g. \cite{Davy99,GPW92,Land03,JMN14,Raja02,Safr94,MS04}).
The possibility of using a linear spectral problem (LSP) to
describe a moving frame (the Gauss--Weingarten equations) on a
surface has yielded many new results concerning the intrinsic
geometry of such surfaces \cite{Bobe94,Gues97,Hele01}. It has
recently proved fruitful to extend this characterization of
soliton surfaces through their immersion functions in Lie
algebras. The construction of such surfaces, related to the
$\mathbb{C}P^{N-1}$ sigma model, has been accomplished by
representing the Euler--Lagrange (E-L) equations as conservation
laws and expressing them in terms of the rank-1 projector
formalism. This allows us to define closed differential 1-forms
for surfaces which can be explicitly integrated
\cite{GY09,Kono96}. This determination has led to a new way of
constructing and investigating two-dimensional surfaces in Lie
algebras, Lie groups and homogeneous spaces
\cite{GG11,FG96,FGFL00,GP11}. The algebraic-geometric approach
based on this formalism applied to the $\mathbb{C}P^{N-1}$ sigma
model equations and associated surfaces has proved to be a
suitable tool for investigating the links between successive
projection operators, wavefunctions of the LSP and immersion
functions of surfaces in the $\mathfrak{su}(N)$ algebra
\cite{GG10}. The main advantages of this approach are that this
formulation preserves the conformal and scaling invariance of
these quantities. It allows us to construct a regular algorithm
for finding certain classes of solutions having a finite action
functional \cite{GS06,GG11}. A broad review of recent developments
in this subject can be found in, e.g. \cite{Zakr89,Hele01,Uhle89}.

In this paper we make use of the fact that the immersion functions
for the surfaces satisfy the same E-L equations as the projectors
which were used to generate the surfaces. The possibility of
further generation: surfaces over surfaces and on, to the whole
stack of surfaces, is analysed. The result is astonishing: the
procedure proves to be idempotent already in its second step. We
next investigate the E-L equations in more detail. If the
constraint of the projective property is imposed on their
solutions they have been considered the equations for rank-1
projectors of the \cpn model. We prove that the class of solutions
which these equations admit is larger, even within the class of
projectors.

In the next section we shortly recall the basic properties of the
\cpn models. Section 3 contains the main results on the
construction of surfaces over surfaces. The theorem on the classes
of projector solutions (not necessarily of rank 1) of the E-L
equations is formulated and proven in Section 4.

\section{Preliminaries on the \cpn models}
 This section
recalls the main notions to be used hereafter, certain theorems
dealing with the rank-1 Hermitian projector analysis and some
techniques for obtaining soliton surfaces via the
$\mathbb{C}P^{N-1}$ sigma models.

In our previous work \cite{GG10,GGP12,GG11} we discussed in detail
the algebraic and analytic properties of 2D-soliton surfaces with
the immersion functions $X_k$, $k=0,1,...,N-1$, which take values
in the $\mathfrak{su}(N)$ algebra and are induced by rank-1
Hermitian projectors $P_k$ of $\mathbb{C}P^{N-1}$ sigma models.
The dynamics of the $\mathbb{C}P^{N-1}$ sigma model defined on the
Riemann sphere $S^2=\mathbb{C}\cup\lbrace\infty\rbrace$ are
determined by stationary points of the action functional
\cite{GG10}
\begin{equation}
\mathcal{A}(P_k)=\int_{S^2}\mbox{tr}(\partial
P_k\bar{\partial}P_k)d\xi d\bar{\xi},\qquad 0\leq k\leq
N-1,\label{1.1}
\end{equation}
where $\partial$ and $\bar{\partial}$ denote the derivatives with respect to $\xi$ and $\bar{\xi}$, respectively
\begin{equation}
\partial=\frac{1}{2}\left(\frac{\partial}{\partial\xi^1}-i\frac{\partial}{\partial\xi^2}\right),\qquad
\bar{\partial}=\frac{1}{2}\left(\frac{\partial}{\partial\xi^1}+i\frac{\partial}{\partial\xi^2}\right),\qquad\xi=\xi^1+i\xi^2.
\end{equation}
The E-L equations within the constraint $P_k^2=P_k$ have the form
\begin{equation}
[\partial\bar{\partial}P_k,P_k]=0.\label{1.3}
\end{equation}
The target space of a projector $P_k$ is determined
by a complex line in $\mathbb{C}^N$, i.e. by a one-dimensional
vector
\begin{equation}
f_k(\xi,\bar{\xi})=\left(f_k^0(\xi,\bar{\xi}),...,f_k^{N-1}(\xi,\bar{\xi})\right)\in\mathbb{C}^N\backslash\lbrace0\rbrace
\end{equation}
related to $P_k$ by
\begin{equation}
P_k=\frac{f_k\otimes f_k^\dagger}{f_k^\dagger f_k},\label{1.5}
\end{equation}
where $\otimes$ denotes the tensor product. In terms of the vector
functions $f_k$, the E-L equations corresponding to the action
functional (\ref{1.1}) take the form \cite{Zakr89}
\begin{equation}\label{f}
\left(\mathbb{I}-\frac{f_k\otimes f_k^\dagger}{f_k^\dagger
f_k}\right)\left[\partial\bar{\partial}f_k-\frac{1}{f_k^\dagger
f_k}\left((f_k^\dagger\bar{\partial}f_k)\partial
f_k+(f_k^\dagger\partial f_k)\bar{\partial}f_k\right)\right]=0,
\end{equation}
where $\mathbb{I}$ is the $N\times N$ identity matrix. Equation
(\ref{1.5}) provides an isomorphism between the equivalence
classes of the $\mathbb{C}P^{N-1}$ model and the set of rank-1
Hermitian projectors $P_k$. An entire class of solutions of
(\ref{f}) is  is obtained by acting on a holomorphic solution
$f_0$ (or antiholomorphic $f_{N-1}$) with raising and lowering
operators \cite{DZ80,Zakr89}.
\begin{eqnarray}\label{rais-lower}
&f_{k+1}=\mathcal{P}_+(f_k)=(\mathbb{I}-P_k)\partial f_k\, ,\nn\\
&f_{k-1}=\mathcal{P}_-(f_k)=(\mathbb{I}-P_k)\bar{\partial} f_k.
\end{eqnarray}
The raising and lowering operators (\ref{rais-lower}) have their
counterparts in the corresponding operators acting on projectors
$P_k$, namely \cite{GG10}
\begin{equation}\label{recurrence}
P_{k+1}=\Pi_+(P_k)=\frac{\partial
P_kP_k\bar{\partial}P_k}{\mbox{tr}(\partial
P_kP_k\bar{\partial}P_k)},\quad
P_{k-1}=\Pi_-(P_k)=\frac{\bar{\partial}P_kP_k\partial
P_k}{\mbox{tr}(\bar{\partial}P_kP_k\partial P_k)}.
\end{equation}
We have proven \cite{GG13} that if $P_k$ satisfies the E-L
equations (\ref{1.3}) then $P_{k+1}$ and $P_{k-1}$ are also
solutions of those equations and the projective property
$P_k^2=P_k$ is preserved by the operations (\ref{recurrence}).

The elements of the set of $N$ rank-1 projectors $\lbrace
P_0,P_1,...,P_{N-1}\rbrace$ satisfy the orthogonality and
completeness relations
\begin{equation}
P_kP_j=\delta_{kj}P_k\quad\mbox{(no summations)},\qquad
\sum_{j=0}^{N-1}P_j=\mathbb{I}.\label{1.8}
\end{equation}
In the papers \cite{GG10,GGP12,GG11,PG12,GSZ05} it was shown that
with each of these solutions we can associate a conformally
parametrized surface in the Lie algebra
$\mathfrak{su}(N)\simeq\mathbb{R}^{N^2-1}$. Namely, the E-L
equations (\ref{1.3}) can be written as the conservation law
\begin{equation}
\partial[\bar{\partial}P_k,P_k]+\bar{\partial}[\partial P_k,P_k]=0.
\end{equation}
This implies that there exist $\mathfrak{su}(N)$ matrix-valued differential 1-forms
\begin{equation}
dX_k=i\left(-[\partial P_k,P_k]d\xi+[\bar{\partial}P_k,P_k]d\bar{\xi}\right),\label{1.10}
\end{equation}
which are closed. For the surfaces corresponding to the rank-1
projectors $P_k$, the integration of (\ref{1.10}) is performed
explicitly \cite{GY09}
\begin{eqnarray}
X_k&=i\int_{\gamma_k}-[\partial P_k,P_k]d\xi+[\bar{\partial}P_k,P_k]d\bar{\xi}\label{1.11a}\\
&=-i\left(P_k+2\sum_{j=0}^{k-1}P_j\right)+ic_k\mathbb{I}\in\mathfrak{su}(N),\qquad
c_k=\frac{1+2k}{N},\label{1.11}
\end{eqnarray}
where $\gamma_k$ is a curve which is locally independent of the
trajectory in $\mathbb{C}$. The $\mathfrak{su}(N)$ immersion
functions $X_k$ satisfy the cubic matrix equations (the minimal
polynomial identity) \cite{GG10,GGP12,GG11}
\begin{equation}
\hspace{-2.5cm}(X_k-ic_k\mathbb{I})(X_k-i(c_k-1)\mathbb{I})(X_k-i(c_k-2)\mathbb{I})=0,\qquad
0<k<N-1\label{1.12}
\end{equation}
for any mixed solution of the E-L equations (\ref{1.3}). For the
holomorphic ($k=0$) or antiholomorphic ($k=N-1$) solutions of the
E-L equations (\ref{1.3}), the minimal polynomial for the
immersion functions $X_0$ and $X_{N-1}$ is quadratic
\begin{equation}
\hspace{-2.5cm}(X_0-ic_0\mathbb{I})(X_0-i(c_0-1)\mathbb{I})=0,\qquad
(X_{N-1}+ic_0\mathbb{I})(X_{N-1}+i(c_0-1)\mathbb{I})=0,\label{1.13}
\end{equation}
(where $c_0+c_{N-1}=2$) and the immersion functions $X_k$ are
linearly dependent
\begin{equation}
\sum_{k=0}^{N-1}(-1)^kX_k=0.
\end{equation}
One of the results obtained in \cite{GGP12} was the derivation of
the E-L equations satisfied by these surfaces, which were
identical to the equations satisfied by the original projectors
$P_k$, namely
\begin{equation}\label{ELX}
[\partial\bar{\partial}X_k,X_k]=0,\qquad X_k^\dagger=-X_k\in\mathfrak{su}(N),
\end{equation}
subject to the constraint (\ref{1.12}),
which encompasses the constraints in (\ref{1.13}). Since the E-L
equations (\ref{1.3}) for the projectors $P_k$ constitute a basis
for the construction of the set of surfaces $X_k$ satisfying the
same E-L equations, a natural question arises as to whether this
technique can be further exploited to construct surfaces over
surfaces and possibly a whole stack of surfaces in a similar way
as the surfaces $X_k$ were built from the projectors $P_k$. The
main goal of this communication is to analyse whether such a
construction is possible. An unexpected result of our analysis is
the fact that these surfaces over surfaces are identical to the
original surfaces up to a multiplicative constant.

\section{Stack of conformally parametrised surfaces}
In this section we prove that the immersion functions of
2D-surfaces $Y_k$ over the surfaces $X_k$, defined analogously to
$X_k$ over the projectors $P_k$, i.e.
\begin{equation}
Y_k(\xi,\bar{\xi})=i\int_{\gamma_k}-[\partial X_k,X_k]d\xi+[\bar{\partial}X_k,X_k]d\bar{\xi}\in\mathfrak{su}(N)\label{2.1}
\end{equation}
are identical to the surfaces $X_k$ up to a multiplication factor
of ($-1$) if we require that the $Y_k$'s be elements of the
$\mathfrak{su}(N)$ algebra. The E-L equations (\ref{1.3}) written
in terms of the $\mathfrak{su}(N)$-valued immersion functions
$X_k$ can be equivalently written as the conservation laws (CLs)
\begin{equation}\label{CL}
\partial [\bar{\partial}X_k,X_k]+\bar{\partial}[\partial X_k,X_k]=0.
\end{equation}
Let us define the $N\times N$ matrix functions
\begin{equation}
M_k:=[\bar{\partial}X_k,X_k],\qquad \mbox{tr}\,M_k=0.\label{2.3}
\end{equation}
Then, we can write the CLs (\ref{CL}) as
\begin{equation}
\partial M_k-\bar{\partial}M_k^\dagger=0.\label{2.4}
\end{equation}
If the CLs (\ref{2.4}) hold, then there
exist matrix-valued differential 1-forms
\begin{equation}
dY_k:=i\left(M_k^\dagger d\xi+M_kd\bar{\xi}\right),\label{2.5}
\end{equation}
which are closed $d(dY_k)=0$ (the factor $i$ was introduced in
order to make the $Y$'s belong to the Lie algebra
$\mathfrak{su}(N)$). From the closedness of the 1-forms
(\ref{2.5}) it follows that the integrated forms of the 2D
surfaces (\ref{2.1}) locally depend only on the end points of the
curves $\gamma_k$ (i.e. they are locally independent of the
trajectory in the complex plane $\mathbb{C}$). The integrals
define mappings $Y_k:\Omega\ni (\xi,\bar{\xi})\mapsto
Y_k(\xi,\bar{\xi})\in\mathfrak{su}(N)\simeq\mathbb{R}^{N^2-1}$,
where $\Omega\in\mathbb{C}$ in an open simply connected domain.
The equation (\ref{2.1}) is known as the generalized Weierstrass
formula for immersion \cite{Kono96}. This gives the following
expressions for the complex tangent vectors
\begin{equation}
\partial Y_k=iM_k^\dagger=-i[\partial X_k,X_k],\qquad \bar{\partial}Y_k=-M_k=i[\bar{\partial}X_k,X_k].
\end{equation}
In the following proofs we are going to use a lemma formulated in our previous work \cite{GG11}.

\begin{lemma}
Let $P_k=P_k(\xi,\bar{\xi}):\mathbb{C}\rightarrow GL(N,\mathbb{C})$ be a rank-1 Hermitian projector determined by a complex line in $\mathbb{C}^N$
\begin{equation}
P_k=\frac{f_k\otimes f_k^\dagger}{f_k^\dagger f_k},\qquad k\in\lbrace0,1,...,N-1\rbrace
\end{equation}
where $f$ is a mapping $\mathbb{C}\supseteq\Omega\ni\xi\mapsto
f=(f_0,f_1,...,f_{N-1})\in\mathbb{C}^N\backslash\lbrace0\rbrace$.
Then the following relations hold for $k\leq N-2$)
\begin{equation}
P_{k+1}\partial P_{k+1}=-\partial P_kP_k,\qquad
\bar{\partial}P_{k+1}P_{k+1}=-P_k\bar{\partial}P_k.\label{24}
\end{equation}
For the proof see \cite{GG11} (equation (25)).
\end{lemma}
We complete the lemma with the two outermost cases, $k=-1$ and $k=N-1$, namely
\begin{equation}
P_0\partial P_0=0,\qquad\mbox{and}\qquad P_{N-1}\bar{\partial}P_{N-1}=0.\label{2.8}
\end{equation}
Proof for the outermost cases:

For any projector $P_k$, the projective property $P_k^2=P_k$ implies
\begin{equation}
P_k\partial P_k=\partial P_k(\mathbb{I}-P_k),\qquad \partial
P_kP_k=(\mathbb{I}-P_k)\partial P_k.\label{2.9}
\end{equation}
The same holds for the $\bar{\partial}$ derivative. We have
\begin{eqnarray}
P_0\partial P_0&=\partial P_0(\mathbb{I}-P_0)=\partial\left(\frac{f_0\otimes f_0^\dagger}{f_0^\dagger f_0}\right)(\mathbb{I}-P_0)\nonumber\\
&=\partial\left(\frac{f_0}{f_0^\dagger f_0}\right)\otimes
f_0^\dagger(\mathbb{I}-P_0)+\frac{f_0}{f_0^\dagger
f_0}\otimes\left(\partial
f_0^\dagger\right)(\mathbb{I}-P_0).\label{28}
\end{eqnarray}
For a holomorphic function $f_0$, its Hermitian conjugate
$f_0^\dagger$ is  antiholomorphic, which means that $\partial
f_0^\dagger=0$. On this basis the second term in (\ref{28})
vanishes, while the first term vanishes due to the orthogonality
of $(\mathbb{I}-P_0)$ to $f_0^\dagger$.

The proof of the second part of (\ref{2.8}) is analogous (where
holomorphic and antiholomorphic are interchanged).\hfill$\square$

Taking the Hermitian conjugates of (\ref{2.8}), we also get
\begin{equation}
\bar{\partial}P_0P_0=0,\qquad \partial P_{N-1}P_{N-1}=0.
\end{equation}

We first demonstrate the usefulness of Lemma 1 by proving equation
(\ref{1.11}) through straightforward induction (previously
proven in \cite{GY09} by a
different method).

\begin{proposition}
Let the functions $X_k$ be $\mathfrak{su}(N)$-valued immersion
functions defined by the differential 1-form (\ref{1.10}) or
equivalently the integral (\ref{1.11a}). Then the immersion
functions $X_k$ can be explicitly written as (\ref{1.11}).
\end{proposition}

Proof. For $k=0$ and from equation (\ref{1.10})
we have
\begin{equation}
\hspace{-2cm}\partial X_0 = -i[\partial P_0,P_0]=-i\left(\partial
P_0P_0-P_0\partial P_0\right)=-i(\mathbb{I}-2P_0)\partial P_0=-i
\partial P_0,
\end{equation}
since $P_0\partial P_0=0$, hence by Hermitian conjugation (which
changes the sign of $X_k$) we have $\bar{\partial} X_0 =
-i\bar\partial{P_0}$. Therefore, the integrated form of the surface is
$X_0=-iP_0+\frac{i}{N}\mathbb{I}$ since the integration constant
$\frac{i}{N}\mathbb{I}$ is unique if we require the matrix $X_0$
to be traceless. Now let
\begin{eqnarray}
\partial X_k&=-i[\partial P_k,P_k]=-i\partial P_k-2i\sum_{j=0}^{k-1}\partial P_j,\label{2.22}\\
\bar{\partial}X_k&=i[\bar{\partial}P_k,P_k]=-i\bar{\partial}P_k-2i\sum_{j=0}^{k-1}\bar{\partial}P_j,
\end{eqnarray}
be satisfied for $k=m, \quad m\in\lbrace 0,...,N-2\rbrace$
(induction hypothesis). We will prove that these equations
(\ref{2.22}) hold for $k=m+1$. We have
\begin{eqnarray}\label{m+1}
[\partial P_{m+1},P_{m+1}]&=-P_{m+1}\partial P_{m+1}+\partial P_{m+1}P_{m+1}\nn\\
&=-P_{m+1}\partial P_{m+1}+(\mathbb{I}-P_{m+1})\partial P_{m+1}\nn\\
&=-2P_{m+1}\partial P_{m+1}+\partial P_{m+1}\nn\\
&=2\partial P_mP_m+\partial P_{m+1},
\end{eqnarray}
where we have used Lemma 1 in the form (\ref{24}) to get
the last line.
%\textsl{
%\begin{equation}
%P_{m+1}\partial P_{m+1}=-\partial P_mP_m.
%\end{equation}}
Furthermore, from (\ref{24}), expression (\ref{m+1}) is equal
\begin{eqnarray}
\partial P_{m+1}+\partial P_mP_m+(\mathbb{I}-P_m)\partial P_m=\partial P_{m+1}+[\partial P_m,P_m]+\partial
P_m.
\end{eqnarray}
Now we replace $[\partial P_m,P_m]$ from the induction
hypothesis ((\ref{2.22}) for k=m). Hence we obtain
\begin{eqnarray}
[\partial P_{m+1},P_{m+1}]&=\partial P_{m+1}+\partial P_m+\partial P_m+2\sum_{j=0}^{m-1}\partial P_j\texttt{\nn}\\
&=\partial P_{m+1}+2\sum_{j=0}^m\partial P_j.\label{2.23}
\end{eqnarray}
Equation (\ref{2.23}) constitutes the induction conclusion. The induction implies that
\begin{equation}
[\partial P_k,P_k]=\partial P_k+2\sum_{j=0}^{k-1}\partial P_j\qquad\mbox{for all }0\leq k\leq N-1.\label{eq-2}
\end{equation}
By Hermitian conjugation (which reverses the order of
multiplication and thus changes the sign of the commutator) we
obtain
\begin{equation}
[\bar{\partial}P_k,P_k]=-\bar{\partial}P_k-2\sum_{j=0}^{k-1}\bar{\partial}P_j.\label{eq-1}
\end{equation}
Integration of (\ref{eq-2}) and (\ref{eq-1}) over the path
$\gamma_k$ yields (\ref{1.11}) if we bear in mind that the
constant of integration is unique to ensure
tracelessness.\hfill$\square$

\begin{corollary}
For each of the complex tangent vectors $\partial X_k$ and $\bar{\partial}X_k$, we equate the two expressions given in (\ref{2.22}) and obtain
\begin{equation}
\partial P_k+2\sum_{j=0}^{k-1}\partial P_j=\partial P_kP_k-P_k\partial P_k=\partial P_k-2P_k\partial
P_k,
\end{equation}
where we have used $\partial P_kP_k=P_k(\mathbb{I}-P_k)$ from
(\ref{2.9}). Thus, we get
\begin{equation}
\sum_{j=0}^{k-1}\partial P_j=-P_k\partial P_k,
\end{equation}
and its respective Hermitian conjugate
\begin{equation}
\sum_{j=0}^{k-1}\bar{\partial}P_j=-\bar{\partial}P_kP_k.
\end{equation}
\end{corollary}

\begin{corollary}
From the orthogonality property of the projectors (\ref{1.8}), we get
\begin{equation}
\partial P_k\sum_{j=0}^{k-1}P_j=-P_k\sum_{j=0}^{k-1}\partial P_j=-P_k(-P_k\partial P_k)=P_k\partial P_k,\label{2.27}
\end{equation}
and
\begin{equation}
\left(\sum_{j=0}^{k-1}P_j\right)\partial P_k=-\left(\sum_{j=0}^{k-1}\partial P_j\right)P_k=P_k\partial P_kP_k=0,\label{2.28}
\end{equation}
together with their respective Hermitian conjugate equations
\begin{equation}
\left(\sum_{j=0}^{k-1}P_j\right)\bar{\partial}P_k=\bar{\partial}P_kP_k,\qquad \bar{\partial}P_k\sum_{j=0}^{k-1}P_j=0.
\end{equation}
\end{corollary}

Under these circumstances we have the following.

\begin{proposition}
Let the $\mathbb{C}P^{N-1}$ model be defined on the Riemann sphere and have a finite action functional. Then the surfaces over surfaces defined by (\ref{2.1}) are identical to the initial surfaces (\ref{1.11}) from which they were derived, up to a factor of ($-1$).
\end{proposition}

Proof. The proof is done by direct calculation from (\ref{2.1})
\begin{eqnarray}
\partial Y_k&=-i[\partial X_k,X_k]=(-i)^3\left[[\partial P_k,P_k],P_k+2\sum_{j=0}^{k-1}P_j\right]\nn\\
%&=i\left[\partial P_kP_k-P_k\partial P_k,P_k+2\sum_{j=0}^{k-1}P_j\right]\nn\\
&=i\left(\partial P_kP_kP_k+2\partial P_kP_k\sum_{j=0}^{k-1}P_j-P_k\partial P_kP_k-2P_k\partial P_k\sum_{j=0}^{k-1}P_j\right.\nn\\
&\hspace{1cm}\left.-P_k\partial P_kP_k-2\sum_{j=0}^{k-1}P_j\partial P_kP_k+P_kP_k\partial P_k+2\sum_{j=0}^{k-1}P_jP_k\partial P_k\right)\nn\\
&=i\left(\partial P_kP_k+0-0-2P_k\partial P_k-0+0+P_k\partial P_k+0\right)\nn\\
&=i(\partial P_kP_k-P_k\partial P_k)=i[\partial P_k,P_k]=-\partial X_k,\label{2.30}
\end{eqnarray}
where we have used Corollary 2 equations (\ref{2.27}) and (\ref{2.28}). The Hermitian conjugate equation is
\begin{equation}
-\bar{\partial}Y_k=-(-\bar{\partial}X_k)=\bar{\partial} X_k.\label{2.31}
\end{equation}
Next, integrating (\ref{2.30}) and (\ref{2.31}) we obtain the vanishing of the expressions
\begin{equation}
X_k+Y_k=0,
\end{equation}
where the constant of integration is chosen to be $-c_k$ in order
to ensure the tracelessness of the immersion functions
$Y_k$, which completes the proof.\hfill$\square$

To summarize, we have provided an explicit expression for 2D
conformally parametrised surfaces induced by surfaces and
demonstrated that these surfaces coincide with the original
surfaces for any recurrence index $k$. This proof demonstrates the
uniqueness of soliton surfaces obtained from the
$\mathbb{C}P^{N-1}$ sigma models. In this way, our attempt to
build the stack in which each next step is a surface over the
previous step, becomes idempotent. This somewhat unexpected result
provides important information on the structure of soliton surface
constructions in the $\mathfrak{su}(N)$ algebra.

\section{Higher-rank projectors as solutions of the Euler-Lagrange
equations}
 It is interesting that the E-L equations (\ref{1.3})
with the projective property $P^2=P$ admit a larger class of
solutions than the rank-1 Hermitian projectors $P_k$.

\begin{proposition}
Let $P$ be a linear combination of rank-1 orthogonal projectors
which satisfy the E-L equations (\ref{1.3})
\begin{equation}
P=\sum_{i=0}^{N-1}\lambda_iP_i,\quad \lambda_i\in\mathbb{C},\qquad
[\partial\bar{\partial} P_i,P_i]=0,
\end{equation}
where not all $\lambda_i$ are zero. Then  P also satisfies the E-L
equations (\ref{1.3}). If, in addition, for all
$i\in\lbrace0,...,N-1\rbrace$ we have $\lambda_i=0$ or
$\lambda_i=1$, then $P$ satisfies $P^2=P.$
\end{proposition}

Proof. We first show that if $P_k$ satisfies the E-L equations
(\ref{1.3}) then its second mixed derivative can be represented as
a combination of at most three rank-1 neighbouring projectors,
namely
\begin{equation}\label{decomp}
\partial\bar{\partial}P_k=\alpha_k P_{k-1}-(\alpha_k+\bar{\alpha}_k) P_k+\bar{\alpha}_k P_{k+1},\label{2.34}
\end{equation}
where
\begin{equation}
\alpha_k=\mbox{tr}(P_k\partial P_k\bar{\partial} P_k).
\end{equation}
For $k=0$ the first component of (\ref{decomp}) vanishes, for
$k=N-1$ the last one does.

We have
\begin{eqnarray}\label{ddP}
(\partial\bar{\partial}P_k)&=\partial(P_k\bar{\partial}
P_k)+\partial[(\mathbb{I}-P_k)\bar{\partial}
P_k]=\partial P_k\bar{\partial} P_k+P_k\partial\bar{\partial} P_k+\partial(\bar{\partial} P_k P_k)\nn\\
&=\partial P_k\bar{\partial} P_k+\bar{\partial} P_k\partial
P_k+2\,\mbox{tr}(P_k\partial\bar{\partial} P_k)P_k
\end{eqnarray}
where we have used the E-L equations (\ref{1.3}), property
(\ref{2.9}) and the obvious fact that for any {rank-1} projector
$P$ and any square matrix $A$ of the same dimension, we have
$PAP=\mbox{tr}(PA)P$ see \cite{GG13}.

Bearing in mind that $\mbox{tr}(P_k\partial P_k)=0$, we may write
the last component of (\ref{ddP}) in terms of $\alpha$ and
$\bar{\alpha}$
\begin{eqnarray}
&2\,\mbox{tr}(P_k\partial\bar{\partial}P_k)P_k =0
-2\,\mbox{tr}(\bar{\partial}P_k\partial
P_k)P_k\nn\\&=2[-\mbox{tr}(P_k\bar{\partial}P_k\partial
P_k)-\mbox{tr}((\mathbb{I}-P_k)\bar{\partial}P_k\partial
P_k)]P_k=-2(\alpha_k+\bar{\alpha}_k)P_k,
\end{eqnarray}
as the projector $(\mathbb{I}-P_k)$ turns into $P_k$ when it
passes $\bar{\partial}P_k$, and the argument of the trace may be
cyclically permuted.

The first component of (\ref{ddP}) can be transformed with the use
of the same property (\ref{2.9}) of rank-1 projectors and with the
recurrence formula (\ref{recurrence})
\begin{eqnarray}\label{dPdP}
\partial P_k\bar{\partial} P_k&=\partial P_k\,P_k\bar{\partial} P_k+\partial P_k (\mathbb{I}-P_k)\bar{\partial} P_k\nn\\
&=\mbox{tr}(\partial P_k\,P_k\bar{\partial}
P_k)P_{k+1}+\mbox{tr}(P_k\partial P_k\bar{\partial}
P_k)P_k=\bar{\alpha}_k P_{k+1}+ \alpha_k P_k
\end{eqnarray}
for $k=0,...,N-2$, while $\partial P_{N-1}P_{N-1}=0$ and
(\ref{dPdP}) reduces to $\alpha_{N-1}P_{N-1}$ for $k=N-1$.\\
 Similarly, the second component of (\ref{ddP}) is given by
\begin{equation}
\bar{\partial} P_k\partial P_k = \alpha_k P_{k-1}+\bar{\alpha}_k
P_k.
\end{equation}
for $k=1,...,N-1$, while $P_0\partial P_0=0$ and (\ref{dPdP})
reduces to $\bar{\alpha}_0P_0$ for $k=0$.

Summing up the three components, we get the required decomposition
of $\partial\bar{\partial}P_k$ (\ref{2.34}).

It follows from (\ref{2.34}) that any linear combination of rank-1
orthogonal projectors $P$ satisfies the E-L equation
\begin{equation}
\hspace{-2.6cm}\left[\partial\bar{\partial}\sum_{i=0}^{N-1}\lambda_iP_i,\sum_{j=0}^{N-1}\lambda_jP_j\right]=\sum_{i,j=0}^{N-1}\lambda_i
\lambda_j\lbrace\alpha_i[P_{i-1},P_j]+\bar{\alpha}_i[P_{i+1},P_j]-\left(\alpha_i+\bar{\alpha}_i\right)[P_i,P_j]\rbrace=0,
\end{equation}
since the projectors $P_i$ are mutually orthogonal (\ref{1.8}). The idempotency condition for the projector $P$ requires that
\begin{eqnarray}
\sum_{i=0}^{N-1}\lambda_iP_i&=P=P^2=\left(\sum_{i=0}^{N-1}\lambda_iP_i\right)^2\\
&=\sum_{i,j=0}^{N-1}\lambda_i\lambda_jP_iP_j=\sum_{i,j=0}^{N-1}\lambda_i\lambda_j\delta_{ij}P_i=\sum_{i=0}^{N-1}\lambda_i^2P_i.\label{2.38}
\end{eqnarray}
In the last equality of (\ref{2.38}) we have again used the
orthogonality property (\ref{1.8}). Hence $\lambda_i^2=\lambda_i$,
which implies that $\lambda_i=0$ or $\lambda_i=1$ for all
$i\in\lbrace 0,...,N-1\rbrace$.\hfill$\square$

Thus Proposition 3 proves that the E-L equations
have a much larger class of solutions
possessing the projective property than
the class of rank-1 projectors $P_k$. This
increases the range of projectors $P$ solvable by the technique
described in this paper.

A similar proposition holds for the immersion function $X$.

\begin{proposition}
Let a function $X\in\mathfrak{su}(N)$ be a linear combination of
immersion functions $X_k$ of 2D-soliton surfaces in the
$\mathfrak{su}(N)$ algebra
\begin{equation}
X=\sum_{k=0}^{N-1}\lambda_k X_k,\qquad \lambda_k\in \mathbb{C},
\end{equation}
where not all $\lambda_k$ are  zero, and the $X_k$ satisfy the E-L
equation (\ref{ELX}). Then
\begin{equation}
[\partial\bar{\partial}X,X]=0\label{2.40}
\end{equation}
holds. If all $\lambda_k$ are real, then the immersion function of
the multileaf surface $X$ is also an element of the
$\mathfrak{su}(N)$ algebra.
\end{proposition}

Proof. Each immersion function $X_k$ is a linear combination, with
constant coefficients, of projectors $P_j$ and the unit matrix
(\ref{1.11}), while, for each $j$, $\partial\bar{\partial}P_j$ is
a linear combination of projectors, given by (\ref{2.34}). Hence
also the mixed second derivative of $X_k$ is a linear combination
of projectors, namely
\begin{equation}
\partial\bar{\partial}X_k=i[\alpha_k
P_{k-1}+(\bar{\alpha}_k-\alpha_k)P_k-\bar{\alpha}_k P_{k+1}].
\end{equation}
Hence $[\partial\bar{\partial}X,X]$ is a linear combination of
commutators, either between projectors or between projectors and
the unit matrix, i.e. all commutators are equal to zero.

If in addition all the $\lambda_k$ are real, it ensures the
anti-Hermitian property of the matrix $X$, while its tracelesness
follows from the fact that it is a linear combination of traceless
matrices. These properties make $X$ an element of
$\mathfrak{su}(N)$. \hfill$\square$

The real part of the coefficients $\alpha_k$ has physical and
geometric interpretations. Namely $\alpha_k+\bar{\alpha}_k =
\mbox{tr}(\partial P_k\bar{\partial} P_k)$ is the Lagrangian
density in the action functional (\ref{1.1}). Moreover, we have
shown in \cite{GGP12} that $\mbox{tr}(\partial X_k\bar{\partial}
X_k)=-\mbox{tr}(\partial P_k\bar{\partial} P_k)$, which makes this
quantity also the Lagrangian density for the surface immersion
functions. It is also the non-diagonal element $g_{12}=g_{21}$ of
the metric tensor on the surface $X_k$, while the diagonal
elements of the metric tensor are zero \cite{GG10}. This way
$\alpha_k+\bar{\alpha}_k$ determines the metric properties of the
surfaces $X_k$ (and obviously all the surfaces of the stack).

To conclude we have shown that the E-L equations (\ref{2.40}) may
also describe multileaf surfaces in addition to the surfaces
generated by rank-1 projectors, which were the subjects of earlier
work \cite{GS06}. This makes soliton surfaces associated with
$\mathbb{C}P^{N-1}$ models a rather special and interesting
subject to study.

\section*{Acknowledgements}
AMG's work was supported by a research grant from NSERC of Canada.
PPG wishes to thank the Centre de Recherches Math\'ematiques
(Universit\'e de Montr\'eal) for the NSERC financial support
provided for his visit to the CRM.


\section*{References}
\begin{thebibliography}{10}
\expandafter\ifx\csname urlstyle\endcsname\relax
  \providecommand{\doi}[1]{doi:\discretionary{}{}{}#1}\else
  \providecommand{\doi}{doi:\discretionary{}{}{}\begingroup
  \urlstyle{rm}\Url}\fi

\bibitem{Bobe94}
Bobenko A.I. (1994) Surfaces in terms of 2 by 2 matrices, Harmonic
Maps and integrable systems. Eds Fordy A. and Wood J.C.,
Braunschweig Vieweg.

\bibitem{Davy99}
Davydov A. (1999) Solitons in Molecular Systems, New York, Kluwer.

 \bibitem{DZ80}
Din A.M. and Zakrzewski W.J. (1980) General classical solutions of
the $\mathbb{C}P^{N-1}$ model, \textit{Nucl. Phys.} B \textbf{174}
397--403.

\bibitem{DHZ84}
Din A.M., Horvath Z. and Zakrzewski W.J. (1984) The
Riemann--Hilbert problem and finite action $\mathbb{C}P^{N-1}$
solutions, \textit{Nucl. Phys. B} \textbf{233} 269--299.

\bibitem{FG96}
Fokas A.S. and Gel'fand I.M. (1996) Surfaces on Lie groups, on Lie
algebras and their integrability, \textit{Commun. Math. Phys.}
\textbf{177} 203--220.

\bibitem{FGFL00}
Fokas A.S., Gel'fand I.M., Finkel F. and Liu Q.H. (2000) A formula
for constructing infinitely many surfaces on Lie algebras and
integrable equations, \textit{Sel. Math.} \textbf{6} 347--375.

\bibitem{GG10}
Goldstein P.P. and Grundland A.M. (2010) Invariant recurrence
relations for $\mathbb{C}P^{N-1}$ models, \textit{J. Phys. A:
Math. Gen.} \textbf{43} 265206.

\bibitem{GG13}
Goldstein P.P. and Grundland A.M. (2011) Invariant description of
$\mathbb{C}P^{N-1}$ sigma models, \textit{Theor. Math. Phys.}
\textbf{168} 939--950.

\bibitem{GG11}
Goldstein P.P. and Grundland A.M. (2011) On the surfaces
associated with $\mathbb{C}P^{N-1}$ models, \textit{J. Phys. C:
Conf. Ser.} \textbf{284} 012031 (9pp).

\bibitem{GGP12}
Goldstein P.P., Grundland A.M. and Post S. (2012) Soliton surfaces
associated with sigma models: differential and algebraic aspects,
\textit{J. Phys. A: Math. Theor.} \textbf{45} 395208 (19pp).

\bibitem{GPW92}
Gross D.G. Piran T. and Weinberg S. (1992) Two Dimensional Qunatum
Gravity and Random Sufaces, Singapore, World Scientific.

\bibitem{GS06}
Grundland A.M. and Snobl L. (2006) Description of surfaces
associated with $\mathbb{C}P^{N-1}$ sigma models on Minkowski
space, \textit{J. Geom. Phys.} \textbf{56} 512--531.

\bibitem{GP11}
Grundland A.M. and Post S. (2011) Soliton surfaces associated with
generalized symmetries of integrable equations, \textit{J. Phys.
A: Math. Theor.} \textbf{44} 165203 (31pp.).

\bibitem{GY09}
Grundland A.M. and Yurdusen I. (2009) On analytic descriptions of
two-dimensional surfaces associated with the $\mathbb{C}P^{N-1}$
sigma models, \textit{J. Phys. A: Math. Gen.} \textbf{42} 172001
(5pp).%%

\bibitem{GSZ05}
Grundland A.M., Strasburger A. and Zakrzewski W.J. (2005) Surfaces
immersed in $\mathfrak{su}(N+1)$ Lie algebras obtained from the
$\mathbb{C}P^{N-1}$ sigma models, \textit{J. Phys. A: Math. Gen.}
\textbf{39} 9187--9213.

\bibitem{Gues97}
Guest M.A. (1997) \textit{Harmonic Maps, Loop Groups and
Integrable Systems} (Cambridge: Cambridge University Press).

\bibitem{Hele01}
Helein F. (2001) Constant Mean Curvature Surfaces, Harmonic Maps
and integrable Systems, Boston, Birkh\"auser.

\bibitem{JMN14}
Jensen G.R., Musso E and Nicolodi L. (2014) The geometric Cauchy
problem for the membrane shape equation, \textit{J. Phys. A: Math.
Theor.} \textbf{47} 495201 (22pp.).

\bibitem{Kono96}
Konopelchenko B. (1996) Induced surfaces and their integrable
dynamics \textit{Stud. Appl. Math.} \textbf{69} 9--51.

\bibitem{Land03}
Landolfi G. (2003) On the Canham--Helfrich membrane model,
\textit{J. Phys. A: Math. Theor.} \textbf{36} 4699 (16pp.).

\bibitem{MS04}
Manton N. and Sutcliffe P. (2004) \textit{Topological Solitons}
(Cambridge: Cambridge University Press).

\bibitem{PG12}
Post S. and Grundland A.M. (2012) Analysis of $\mathbb{C}P^{N-1}$
sigma models via projective structures, \textit{Nonlinearity}
\textbf{25} (36pp).

\bibitem{Raja02}
Rajaraman R. (2002) $\mathbb{C}P^n$ solitons in quantum Hall
systems, \textit{Eur. Phys. J. B} \textbf{28} 157--162.

\bibitem{Safr94}
Safran S. (1994) Statistical Thermodynamics of surfaces Interface
and Membranes, Massachusetts, Addison--Wesley.

\bibitem{Uhle89}
Uhlenbeck K. (1989) Harmonic maps into Lie groups (classical
solutions of the chiral model) \textit{J. Diff. Geom.} \textbf{30}
1--50.

\bibitem{Zakr89}
Zakrzewski W.J. (1989) \textit{Low Dimensional Sigma Models}
(Bristol: Adam Hilger 1989) Chapter 3 (46--74).

%\bibitem{Gues97}
%Guest M.A. 1997 Harmonic Maps, Loop Groups and Integrable systems, Cambridge, Cambridge Univ. Press.


\end{thebibliography}
\end{document}